\documentclass[twocolumn]{webofc}

\usepackage[varg]{txfonts}   
\usepackage{hyperref}
\usepackage{url}
\hypersetup{colorlinks=true,citecolor=blue,urlcolor=blue,linkcolor=blue}

\usepackage{upgreek}
\usepackage{wasysym}

\newcommand{\be}{\begin{equation}}
\newcommand{\ee}{\end{equation}}
\newcommand{\ba}{\begin{align}}
\newcommand{\ea}{\end{align}}
\newcommand{\bea}{\begin{eqnarray}}
\newcommand{\eea}{\end{eqnarray}}
\newcommand{\nn}{\nonumber}

\newcommand{\amuSM}{\ensuremath{a_\upmu^\text{SM}}}

\newcommand{\amuexp}{\ensuremath{a_\upmu^\text{exp}}}

\newcommand{\amuexpresult}{116\, 592\, 071.5(14.5)}

\newcommand{\amuHVPLOresult}{7132(61)}

\newcommand{\amuHVPNLOresult}{-99.6(1.3)}

\newcommand{\amuHVPNNLOresult}{12.4(1)}
\newcommand{\amuHVPtotalresult}{7045(61)}

\newcommand{\amuHLbLdataresult}{103.3(8.8)}
\newcommand{\amuHLbLlatticeresult}{122.5(9.0)}
\newcommand{\amuHLbLNLOdataresult}{2.6(6)}
\newcommand{\amuHLbLaverageresult}{112.6(9.6)}
\newcommand{\amuHLbLtotalresult}{115.5(9.9)}

\newcommand{\amudiffresult}{38(63)}

\newcommand{\WPold}{Aoyama:2020ynm}
\newcommand{\WPnew}{Aliberti:2025beg}
\newcommand{\HSZ}{Hoferichter:2024bae}
\newcommand{\LMR}{Leutgeb:2022lqw}
\newcommand{\Tref}{Cappiello:2025fyf,Mager:2025pvz}

\newcommand{\HLbLref}{Colangelo:2015ama,Masjuan:2017tvw,Colangelo:2017fiz,Hoferichter:2018kwz,Eichmann:2019tjk,Bijnens:2019ghy,Leutgeb:2019gbz,Cappiello:2019hwh,Masjuan:2020jsf,Bijnens:2020xnl,Bijnens:2021jqo,Danilkin:2021icn,Stamen:2022uqh,Leutgeb:2022lqw,Hoferichter:2023tgp,Hoferichter:2024fsj,Estrada:2024cfy,Ludtke:2024ase,Deineka:2024mzt,Eichmann:2024glq,Bijnens:2024jgh,Hoferichter:2024bae,Holz:2024diw,Cappiello:2025fyf,Mager:2025pvz,Colangelo:2014qya,Blum:2019ugy,Chao:2021tvp,Chao:2022xzg,Blum:2023vlm,Fodor:2024jyn}

\begin{document}
\title{Status of the hadronic light-by-light contribution to the
muon $g-2$\\ and holographic QCD predictions}
%
%

\author{        
\firstname{Anton} \lastname{Rebhan}\inst{1}\fnsep\thanks{\email{anton.rebhan@tuwien.ac.at}}
\and
\firstname{Luigi} \lastname{Cappiello}\inst{2}\fnsep\thanks{\email{luigi.cappiello@unina.it}}
\firstname{Josef} \lastname{Leutgeb}\inst{1}\fnsep\thanks{\email{josef.leutgeb@tuwien.ac.at}} 
\and
        \firstname{Jonas} \lastname{Mager}\inst{1}\fnsep\thanks{\email{jonas.mager@tuwien.ac.at}}
        }

\institute{Institut f\"ur Theoretische Physik, Technische Universit\"at Wien,
        Wiedner Hauptstrasse 8-10, A-1040 Vienna, Austria \label{addr1}
        \and
        Dipartimento di Fisica, Universit\`a di Napoli ``Federico II", and
INFN-Sezione di Napoli, Via Cintia, I-80126 Napoli, Italy}


\abstract{%
  We review the recent progress made with regard to the hadronic light-by-light (HLbL)
  contribution to the Standard Model prediction of the muon anomalous magnetic moment
  and how well this compares with predictions from holographic QCD models, which
  had predicted larger contributions from axial vector mesons and short-distance
  constraints than the White Paper of 2020. A new holographic prediction concerns
  tensor-meson contributions, which in holographic QCD play a significant role
  in short-distance constraints beyond the Melnikov-Vainshtein constraint.
  When matching also the symmetric longitudinal short-distance constraint, the
  resulting prediction for the tensor-meson transition form factors agree
  well with available singly virtual data, but lead to different results
  than the traditional quark-model ansatz and a sizable positive contribution
  that could explain the remaining current tension between lattice and data-driven results for the HLbL contribution.
}
\maketitle
\fancyhead[RO,LE]{\thepage}

\section{Introduction}

With the successful completion of the Muon $g-2$ Experiment at the Fermi National
Accelerator Laboratory in June 2025, the magnetic anomaly $a_\upmu=(g-2)_\upmu/2$
is now known with a precision of 124ppb when all available experimental data
are combined, namely
\cite{Muong-2:2025xyk}
\begin{equation}
\label{eq:amuexpresult}
    \amuexp=\amuexpresult\times 10^{-11}\,.
\end{equation}
Shortly before, the international Muon $g-2$ Theory Initiative has released
an update \cite{\WPnew} (WP25) of the Standard Model (SM) prediction.
In contrast to the WP20 result \cite{\WPold} the new one completely agrees
with the experimental value within errors, owing to the exclusive use of
lattice-QCD evaluations of the hadronic vacuum polarization (HVP)
as new experimental data on the
hadronic cross-section $e^+ e^-\to$ hadrons are contradicting
those employed previously to derive HVP from a dispersion integral.

Combined with a new evaluation of the hadronic light-by-light (HLbL) contribution
\cite{\WPnew,\HLbLref}
where the theoretical uncertainty could be reduced from
$18\times 10^{-11}$ to $9.9 \times 10^{-11}$
the new SM prediction 
and the experimental world average no longer show any tension at the
current (somewhat lower) level of precision ($63\times 10^{-11}$ instead of the $43\times 10^{-11}$ estimated in
the WP20 \cite{\WPold}):
\begin{equation}
\label{amudiff}
 \Delta a_\upmu\equiv\amuexp - \amuSM =\amudiffresult\times 10^{-11}\,.
\end{equation}
Table \ref{tab:summary_comparison} displays the status of the hadronic
contributions, which are solely responsible for the final
error budget, and their development since WP20.

\begin{table}[b]
	\caption{Comparison of hadronic contributions according to WP25 \cite{Aliberti:2025beg} to the corresponding numbers from WP20~\cite{Aoyama:2020ynm}} 
\small
\centering
	\begin{tabular}{lrr}
	\hline
	   Hadronic contribution ($\times 10^{11}$)  & WP25 & WP20\\ 
       \hline
HVP LO (lattice) &  $\amuHVPLOresult$ & $7116(184)$\\
HVP LO ($e^+e^-,\tau$) &  \lightning & $6931(40)$ \\
HVP NLO ($e^+e^-$) & $\amuHVPNLOresult$ & $-98.3(7)$\\
HVP NNLO ($e^+e^-$) & $\amuHVPNNLOresult$ & $12.4(1)$\\
HLbL (phenomenology) &  $\amuHLbLdataresult$ & $92(19)$\\
HLbL NLO (pheno) &   $\amuHLbLNLOdataresult$ & $2(1)$\\
HLbL (lattice) &  $\amuHLbLlatticeresult$ &  $82(35)$\\
HLbL (pheno + lattice) &  $\amuHLbLaverageresult$ & $90(17)$\\
\hline
HVP (LO + NLO + NNLO)& $\amuHVPtotalresult$ & $6845(40)$ \\
HLbL 
&   $\amuHLbLtotalresult$ & $92(18)$\\ 
        \hline
        \renewcommand{\arraystretch}{1.0}
	\end{tabular}
\label{tab:summary_comparison}
\end{table}

The anomalous magnetic moment of the muon will continue to serve as
an important tool in the quest for physics beyond the SM, and this
tool needs to be sharpened above all by resolving the current
contradictions between different experiments and lattice simulations
for HVP. There is reasonable hope that a substantially improved
SM prediction will become available in due course.

The second most important theoretical error from HLbL scattering appears to be sufficiently under control already, but
it is not much below the experimental error and thus
will need further work as well eventually. In fact,
the current White Paper result for the HLbL contribution involves a scale factor of 1.5 due to a certain tension between analytic and lattice approaches; hence, a correspondingly smaller HLbL error could be obtained by resolving this issue.

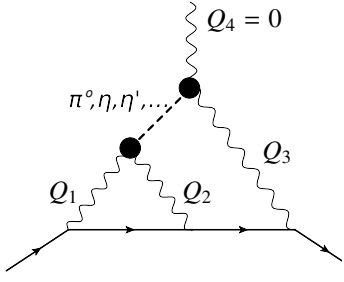
\begin{figure}[t]
\medskip
\begin{center}
\includegraphics[width=0.275\textwidth,clip]{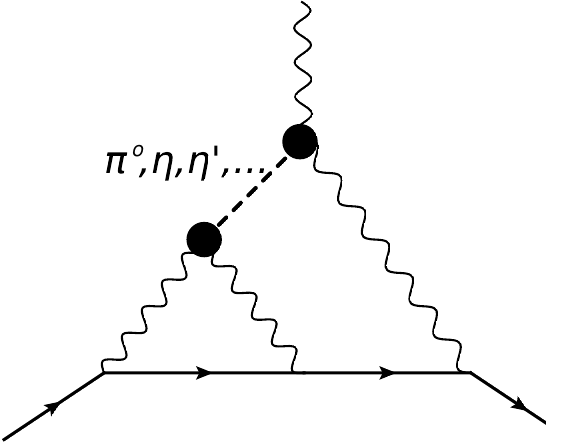}
\end{center}
\begin{picture}(0,0)
 \put(66,50){$Q_1$}
 \put(116,50){$Q_2$}
 \put(146,65){$Q_3$}
 \put(125,115){$Q_4=0$}
\end{picture}
\vspace*{-7mm}
\caption{Hadronic light-by-light scattering contribution to $a_\upmu$ from single meson exchange}
\label{fig:singlemesonexchange}       
\end{figure}

In this contribution, we shall review the progress that has been
achieved between WP20 and WP25 regarding the HLbL contributions, 
with a special emphasis on
checks on the data-driven approach that have been provided
by holographic QCD (hQCD) models. The long-standing open issue on the
size of contributions from axial-vector mesons in single exchange
diagrams like Fig.~\ref{fig:singlemesonexchange} and
related short-distance constraints (SDCs) has now been
tackled by the dispersive approach \cite{\HSZ} in remarkable
agreement with the previous holographic QCD results of \cite{\LMR}.

Ref.~\cite{\HSZ} has also revisited the question of tensor-meson
contributions with the conclusion that they have been underestimated
in previous evaluations \cite{Danilkin:2016hnh}, and moreover that they
contributed with a negative sign.
In both, \cite{\HSZ} and \cite{Danilkin:2016hnh}, a simple quark
model had been assumed. However, in the recent hQCD study
by the present authors \cite{\Tref} even larger and positive contributions have
been obtained, which in fact could resolve the current
tension between the analytic and lattice results on HLbL.

In the following, we review the holographic results for single-meson
exchange contributions and the resulting short-distance behavior of the 
HLbL amplitude
obtained
in \cite{\LMR,\Tref}. Besides the crucial role of the infinite tower
of axial-vector mesons in hQCD, also the one of tensor mesons turns out
to play a significant role. Finally we compare the numerical predictions
of hQCD models with the current data-driven results
of \cite{Hoferichter:2018dmo,Hoferichter:2018kwz,Ludtke:2024ase,Holz:2024lom,Hoferichter:2024bae,Holz:2024diw}. 





\section{Holographic QCD models and SDCs}

The hQCD models considered in \cite{\LMR,\Tref} and in the WP25\cite{Aliberti:2025beg}
are minimal hard-wall (HW) models \cite{Erlich:2005qh,Katz:2007tf},\footnote{The
alternative soft-wall (SW) model \cite{Karch:2006pv} and
its extension studied in \cite{Colangelo:2023een,Colangelo:2024xfh} have been discarded
in the WP25 due to the problems discussed in \cite{Leutgeb:2025jmv}.}
where correlation functions of
left and right-handed quark currents are described holographically
by five-dimensional flavor gauge fields with Yang-Mills action
\bea\label{S5YM}
S_{\mathrm{YM}} &=& -\frac{1}{4g_5^2} \int d^4x \int_0^{z_0} dz
\sqrt{g}\, g^{PR}g^{QS} 
\nn\\
&&\times\;\text{tr}\left(\mathcal{F}^\mathrm{L}_{PQ}\mathcal{F}^\mathrm{L}_{RS}
+\mathcal{F}^\mathrm{R}_{PQ}\mathcal{F}^\mathrm{R}_{RS}\right),
\eea
where $P,Q,R,S=0,\dots,3,z$ and $\mathcal{F}_{MN}=\partial_M \mathcal{B}_N-\partial_N \mathcal{B}_M-i[\mathcal{B}_M,\mathcal{B}_N]$,
in an AdS background geometry with metric
\be\label{ds2AdS}
ds^2=z^{-2}(\eta_{\mu\nu}dx^\mu dx^\nu - dz^2),
\ee
conformal boundary at $z=0$, and a sharp cutoff at $z=z_0 \sim \Lambda^{-1}_\mathrm{QCD}$.
Flavor anomalies are encoded by the Chern-Simons action $S_{\rm CS}=S_{\rm CS}^\mathrm{L}-S_{\rm CS}^\mathrm{R}$ with (in
differential form notation)
\be\label{SCS}
S_{\mathrm{CS}}^\mathrm{L,R}=\frac{N_c}{24\pi^2}\int\text{tr}\left(\mathcal{B}\mathcal{F}^2-\frac{i}2 \mathcal{B}^3\mathcal{F}
-\frac1{10}\mathcal{B}^5\right)^\mathrm{L,R}.
\ee
The $U(1)_A$ anomaly was studied in two different implementations,
the Katz-Schwartz model \cite{Katz:2007tf} in \cite{\LMR} and an implementation
through a scalar-extended Chern-Simons action involving
so-called superconnections in \cite{Leutgeb:2024rfs}.

Scalar and pseudoscalar quark bilinear operators are dual to
a bifundamental scalar field $X_{ij}$ with action
\be
S_{X}=\int d^4x \int_0^{z_0} dz\,\sqrt{g}\;\text{tr}\left(
|DX|^2-M_X^2|X|^2 \right),
\ee
where $D_M X=\partial_M X-i \mathcal{B}_M^\mathrm{L} X+iX\mathcal{B}_M^\mathrm{R}$
and $M_X^2=-3$ as dictated by the scaling dimension of $q\bar q$. The
vacuum solutions in AdS,
$
X(z)=
\frac12(M_q z+\Sigma\, z^3),
$
encode the quark mass matrix $M_q$ and the chiral condensate $\Sigma$.

Together with the Witten-Veneziano mass term associated with
the $U(1)_A$ anomaly this permits to model the nonet of pseudoscalars
with good agreement with regard to their spectrum as well as their two-photon couplings
\cite{\LMR,Leutgeb:2024rfs}. In this setup, the pseudoscalars $\eta$ and $\eta'$ appear
in mixture with a pseudoscalar glueball state.

\subsection{SDCs on TFFs}

Vector meson dominance (VMD) is naturally implemented in hQCD, with an infinite
tower of (degenerate) $\rho,\omega,\Phi$ mesons. Crucially, this
leads to transition form factors (TFFs) for pseudoscalars \cite{Grigoryan:2007wn} and axial-vector
mesons \cite{Leutgeb:2019gbz} which precisely reproduce
the results obtained in light-cone expansions (LCE) \cite{Brodsky:1981rp,Hoferichter:2020lap},\footnote{In fact, the asymptotic form
of the axial-vector TFF was first obtained in hQCD \cite{Leutgeb:2019gbz}
and later found to agree with the LCE result of \cite{Hoferichter:2020lap}.}
whereas VMD ans\"atze with a finite number of vector meson resonances
cannot reproduce the logarithmic functions of the asymmetry parameter $w=(Q_1^2-Q_2^2)/(Q_1^2+Q_2^2)$ with photon virtualities $Q_i$. With $Q^2=(Q_1^2+Q_2^2)/2$
those read
\be
                F_{\pi^0\gamma^*\gamma^*}(Q_1^2,Q_2^2)\to\frac{2 f_\pi}{Q^2}\left[ \frac1{w^2}-\frac{1-w^2}{2w^3}\ln\frac{1+w}{1-w} \right]
                \label{pionTFFas}
\ee
and (in the notation of \cite{Leutgeb:2019gbz})
\begin{align}
                &A_n(Q_1^2,Q_2^2) \to \nonumber\\
                &\frac{12\pi^2 F^A_{n}}{N_c Q^4}
                \frac1{w^4}\left[
                w(3-2w)+\frac12 (w+3)(1-w)\ln\frac{1-w}{1+w}
                \right].
\end{align}

\subsection{Melnikov-Vainshtein SDC}

In \cite{Leutgeb:2019gbz,Cappiello:2019hwh} it was moreover shown
that the infinite tower of axial-vector mesons that is naturally present in
the hQCD models is responsible for reproducing the
Melnikov-Vainshtein SDC \cite{Melnikov:2003xd} of the HLBL scattering amplitude
following from the nonrenormalization theorem for the chiral anomaly and
OPE. In terms of the tensor basis used in \cite{Colangelo:2015ama} it reads
\begin{equation}
\label{eq:MVConstr}
    \lim_{Q_3\rightarrow \infty}\lim_{Q \rightarrow \infty}  Q_3^2 Q^2 \bar{\Pi}_1(Q,Q,Q_3)= -\frac{2}{3 \pi^2}.
\end{equation}
Models with only a finite number of mesons cannot reproduce this, since
each single meson exchange gives zero in this limit, as the propagator in fig.~\ref{fig:singlemesonexchange} goes like $1/Q_3^2$ and the two
form factors fall off as $1/Q^2$ and $1/Q_3^2$.

\begin{figure*}[t]
\caption{The hQCD results for the pion TFF of Ref.~\cite{Leutgeb:2022lqw} (LMR22) with
OPE fit  (blue lines) and  $F_\rho$-fit (red lines) compared to the dispersive results of
Ref.~\cite{Hoferichter:2018kwz} and experimental data. (Figure taken from \cite{\WPnew})}
\label{fig:pi0TFF}       
\centering
\includegraphics[width=0.5\textwidth,clip]{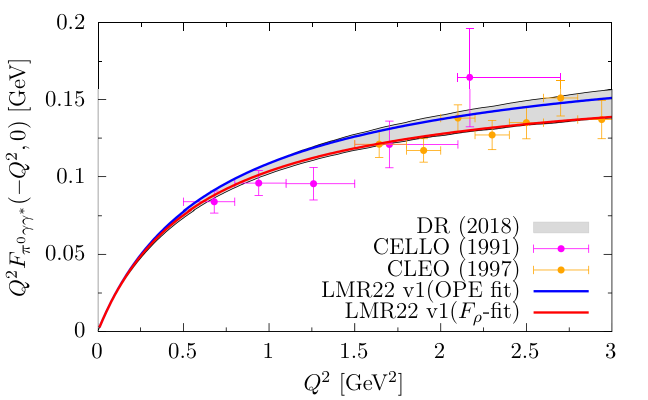}\includegraphics[width=0.5\textwidth,clip]{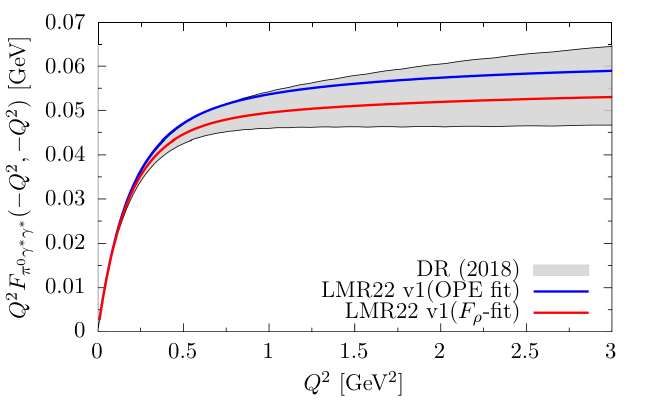}
\end{figure*}

In \cite{Colangelo:2019lpu,Colangelo:2019uex} a Regge model with an infinite
tower of excited pseudoscalars was constructed 
to estimate the effects of the Melnikov-Vainshtein SDC.
However, as in fact noted in \cite{Colangelo:2019uex}, excited pseudoscalar states decouple in the chiral large-$N_c$ limit,
which is not the case for the infinite tower of axial vector mesons. 
The hQCD models of \cite{Leutgeb:2019gbz,Cappiello:2019hwh} thus
provided the first hadronic models that respected the Melnikov-Vainshtein SDC
also in the chiral limit.

\subsection{Symmetric SDCs}

The symmetric short-distance limit of $\bar\Pi_1$, on the other hand, reads
\cite{Melnikov:2003xd,Colangelo:2019lpu,Colangelo:2019uex,Bijnens:2020xnl,Bijnens:2021jqo}
\begin{align}\label{symlongSDC}
    \lim_{Q\to\infty}
    Q^4 \bar\Pi_1(Q,Q,Q)=-\frac4{9\pi^2},
\end{align}
which is not saturated by the tower of axial-vector mesons. The latter only
yield 81.22\% of this result.

In \cite{\Tref} it was shown by the present authors that the infinite tower of
tensor mesons in hQCD models dual to metric fluctuations
\cite{Katz:2005ir} can fill the gap, while not changing
the result for the asymmetric Melnikov-Vainshtein SDC.

Choosing the normalization of tensor modes, which is an extra free parameter
in the model,
such that \eqref{symlongSDC} is reproduced exactly leads to
tensor TFFs in remarkable numerical agreement with
experimental constraints.

Moreover, also the transverse short-distance behavior of the remaining
components of the HLbL tensor, for which the axial-vector meson tower
provides only between 37.5 to 75\% in the symmetric limit,
is raised by the addition of the tensor tower
to around 100\% apart from $\bar\Pi_{8,9}$ where tensor mesons
contribute very little, see Table \ref{tab:symmSDCs} \cite{CLMR26}.

\begin{table}[h]
\caption{
Symmetric $Q\to\infty$ limit results for axial (A) and tensor (T)
contributions to $Q^4\bar\Pi_{1...4}$ and $Q^6\bar{\Pi}_{\ge5}$
divided by the respective quark loop (QL) results in percent.
\label{tab:symmSDCs}
}
\centering
\begin{tabular}{cccc}
\hline
$i$ &  A/QL & T/QL & sum[\%] \\
\hline
1,2 & 81.2 & 18.8 & 100 \\
3,4 & 37.5 & 70.4 & 108 \\
5,6,7 & 75.0 & 31.6 & 107 \\
8,9 & 59.0 & 3.3 & 62 \\
10 & 68.8 & 28.9 & 98 \\
\hline
\end{tabular}
\end{table}

\section{HLbL contributions to $a_\upmu$}

As a model of QCD in a large-$N_c$ expansion, hQCD provides
the HLbL scattering amplitude through a Witten diagram
involving bulk-to-boundary propagators for photons and bulk-to-bulk
propagators for all 5-dimensional fields. It can be represented
as a series of
terms composed of TFFs for (virtual) photons connected
by single-meson propagators in the narrow-width approximation.
The TFFs for the ground-state mesons can be compared with experimental
data and
results obtained in the dispersive approach, permitting
a validation of a given hQCD model before using it to
predict the complete HLbL contribution to $a_\upmu$.

\subsection{Pseudoscalars}

In the HW model considered in \cite{Leutgeb:2022lqw} there are just
enough parameters to fix the mass of the $\rho$ meson,
the pion decay constant $f_\pi$ and the masses of the
nonet of pseudoscalars. The 5-dimensional Yang-Mills
coupling can be determined by matching the short-distance
behavior of the two-point vector correlator to the leading
OPE result of QCD, yielding \cite{Erlich:2005qh} $g_5=2\pi$ for $N_c=3$.
As alternative, \cite{Leutgeb:2021mpu,Leutgeb:2022lqw}
has proposed to fit the $\rho$ meson decay constant,
leading to the somewhat reduced value $g_5=0.894 \times (2\pi)$.
At moderately large virtualities, this leads to a reduction
of the magnitude of TFFs corresponding roughly to the gluonic
corrections obtained in \cite{Melic:2002ij,Bijnens:2021jqo};
this also improves the otherwise deficient
result for HVP in the HW model \cite{Leutgeb:2022cvg}.

Fig.~\ref{fig:pi0TFF} compares the resulting TFFs for $\pi^0$ for the
two choices of $g_5$ with experimental data and the dispersive result of \cite{Hoferichter:2018kwz}, with nearly perfect agreement.
Correspondingly, the hQCD result 
\be
a_\upmu^{\pi^0}(\text{HW}|F_\rho\text{-fit})=63.4\times 10^{-11}
\ee
is close to the central value of the
dispersive result adopted by the WP25, see
Table \ref{tab:amuPS}.

Whereas the hQCD results for $\pi^0$ are rather stable
under variations of the HW model \cite{Leutgeb:2021mpu},
the results for $\eta$ and $\eta'$ depend on the specific
implementation of the U(1)$_A$ anomaly and whether the CS term
includes the bifundamental scalar in form of a superconnection
\cite{Leutgeb:2024rfs}. The ``best-guess'' model of \cite{Leutgeb:2022lqw}, v1($F_\rho$-fit), turns out to predict
somewhat larger values for the TFFs and $a_\upmu$.

\begin{table}[h]
   \caption{Dispersive results for the ground-state pseudoscalar exchange contributions
   used in the White Paper 2025 \cite{\WPnew} compared to the hQCD results
   obtained in the hQCD model v1($F_\rho$-fit) of Ref.~\cite{Leutgeb:2022lqw}, in units of $10^{-11}$}
    \label{tab:amuPS}
    \centering
    \begin{tabular}{ccc}
\hline
 & $a_\upmu^\mathrm{disp.}$ \cite{Hoferichter:2018dmo,Hoferichter:2018kwz,Holz:2024lom,Holz:2024diw} &  $a_\upmu^\mathrm{hQCD}$  \cite{Leutgeb:2022lqw}   \\
 \hline
 $\pi^0$ & $63.0^{+2.7}_{-2.1}$ &  $63.4$  \\
 $\eta$ & $14.7(9)$ &
  $17.6$  \\
 $\eta^\prime$ & $13.5(7)$ &
  $14.9$ \\
 \hline
 Sum & $91.2^{+2.9}_{-2.4}$  & $95.9$\\
 \hline
\end{tabular}
 \end{table}


\subsection{Axial-vector mesons}

\begin{figure}[b]
\centering
\includegraphics[width=0.5\textwidth,clip]{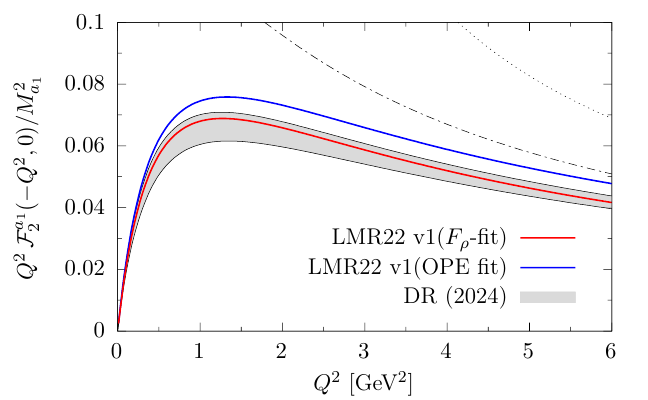}
\caption{The hQCD results for the $a_1$ TFF of Ref.~\cite{Leutgeb:2022lqw} (LMR22) with
OPE fit  (blue lines) and $F_\rho$-fit (red lines) compared to the dispersive result of
Ref.~\cite{Ludtke:2024ase}. Dotted (and dash-dotted) lines show
the LCE limit~\cite{Hoferichter:2020lap}  without (and with) mass corrections. (Figure taken from \cite{\WPnew})}
\label{fig:a1TFF}       
\end{figure}

As is the case for $\pi^0$, the HW hQCD prediction for the
contribution of the axial-vector meson $a_1$ is rather
insensitive to details of the model \cite{Leutgeb:2021mpu} and
close to the chiral HW1 result obtained originally in \cite{Leutgeb:2019gbz}. Rather recently, the first
dispersive result for the singly virtual TFF has been
produced in \cite{Ludtke:2024ase}. As can be seen in
Fig.~\ref{fig:a1TFF}, it fully validates the HW result,
which in the $F_\rho$-fit version is within the error band
of the dispersive result in the entire range of $Q^2$.

Correspondingly the resulting contribution to $a_\upmu$
agrees perfectly with the latest dispersive analysis of \cite{\HSZ},
see Table \ref{tab:HSZIR} which lists the contributions
from photon virtualities $Q_i<1.5$ GeV.
The dispersive results turn out to agree also with the
LMR22 v1($F_\rho$-fit) \cite{Leutgeb:2022cvg} results for the combined contribution of $f_1$ and $f_1'$ as well as for the entire sum
of subleading contributions from the axial sector. In the
dispersive result, the contribution from higher states
are accounted for by effective poles, whereas the hQCD result
involves specific contributions from all the excited axial-vector mesons as well as excited pseudoscalars.

\begin{table}[h] 
\caption{Subleading contributions to $a_\upmu$ in the IR (all $Q_i<1.5$ GeV) from axial-vector mesons, excited axials ($AV^*$),
excited pseudoscalars ($PS^*$), the latter alternatively described by effective poles,
in the dispersive analysis of \cite{\HSZ} in comparison with the hQCD model v1($F_\rho$-fit) of Ref.~\cite{Leutgeb:2022lqw}, in units of $10^{-11}$
}
\label{tab:HSZIR}
    \centering
\begin{tabular}{ccc}
\hline
 & $a_\upmu^\mathrm{disp.,IR}$ \cite{\HSZ} &  $a_\upmu^\mathrm{hQCD,IR}$ \cite{Leutgeb:2022lqw}    \\
\hline
$a_1$ & 3.8(7) & 4.2  \\
$f_1+f_1'$ & 8.4(1.4) & 8.9  \\
$AV^*$  & & 0.7 \\
$PS^*$  & & 1.7 \\
eff.poles &   2.0 & \\
\hline
Sum &  14.2(1.6) & 15.4 \\
\hline
\end{tabular}
\end{table}

\subsection{Tensor mesons}

As already mentioned, the recent dispersive analysis of
subleading contributions \cite{\HSZ} also
included tensor mesons, but only in the form of
a simple quark model with a single TFF (out of five).
The same model had been used before in \cite{Danilkin:2016hnh},
but in a formalism affected by unresolved kinematic
singularities.

\begin{figure}[b]
\caption{Comparison of hQCD result (HW) and quark model
ansatz with $\Lambda=M_\rho$~\cite{Hoferichter:2024bae} and $M_T$~\cite{Hoferichter:2020lap,Schuler:1997yw} for the singly-virtual tensor TFF for helicity $\lambda=2$ with Belle data~\cite{Belle:2015oin} for the $f_2(1270)$.
(Figure adapted from Ref.~\cite{Cappiello:2025fyf})}
\label{fig:f2TFF}       
\centering
\includegraphics[width=0.48\textwidth,clip]{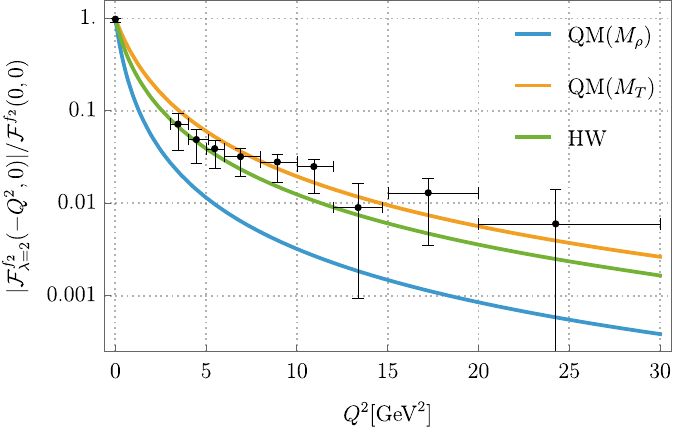}
\end{figure}

In hQCD, the 5-dimensional metric field is dual to the
energy-momentum tensor and was proposed in \cite{Katz:2005ir}
to represent not only a tensor glueball but all the tensor mesons.
In \cite{\Tref}, the present authors have found that
fixing the normalization of tensor mesons such that the
symmetric SDC \eqref{symlongSDC} is saturated
by the combined towers of excited axial-vector and tensor mesons
leads to an agreement of both magnitude and form of
the experimentally measured singly virtual TFF with
helicity 2 for the $f_2(1270)$ meson, see Fig.~\ref{fig:f2TFF}.

However, in contrast to the quark model, the hQCD model
requires a second TFF, conventionally
called $\mathcal{F}_3^T$, which only contributes
in the doubly virtual case. A comparison with resonance chiral
theory indeed suggests that $\mathcal{F}_1^T$ and $\mathcal{F}_3^T$
may be equally important at low energies (see
Appendix A in \cite{Cappiello:2025fyf}). As also found
in \cite{Estrada:2025bty}, $\mathcal{F}_3^T$
can change the sign of the ground-state tensor pole contribution to $a_\upmu$,
and in hQCD it indeed does. When the complete tensor
contribution is considered, the contribution in fact
rises from 2.9 to $8.5 \times 10^{-11}$ in the
region $Q_i<1.5$ GeV, see Table \ref{tab:Qregions},
to be contrasted with $-2.5\times 10^{-11}$ from the quark model adopted in \cite{\HSZ}.

Currently there are no experimental data which could provide
a check on the hQCD prediction for $\mathcal{F}_3^T$.
There is, however, a tension between the current
phenomenological result and the lattice result for $a_\upmu^\mathrm{HLbL}$ (see Table \ref{tab:summary_comparison}) which would be removed by
an extra contribution from the tensor sector of
about $11\times 10^{-11}$ as predicted 
universally by the HW hQCD models with AdS geometry.

\begin{table}[t]
\caption{Subleading contributions in units of $10^{-11}$ from different virtuality regions. HSZ refers to the analysis of 
\cite{\HSZ} using a dispersive approach
for ground-state
axial-vector mesons ($A$) and
scalar mesons $f_0(1370),a_0(1450)$ ($S$), with a
simple quark-model ansatz for
ground-state tensor mesons ($T$),
with short-distance contributions modeled by effective poles;
hQCD lists the corresponding results in the 
model v1($F_\rho$-fit) of \cite{\LMR}
for the axial sector and of 
\cite{Cappiello:2025fyf}
for the tensor sector (marked by a star).
The contributions from ground-state and excited axial-vector states (the
latter in the line ``Other'') agree remarkably well with the HSZ results,
while the tensor sector differs strongly from the quark-model result adopted in HSZ.}
\label{tab:Qregions}
\small
\begin{center}
\begin{tabular}{llccc}
\hline
Region & & HSZ\cite{Hoferichter:2024bae} 
& hQCD (\cite{Leutgeb:2022lqw}+\cite{Cappiello:2025fyf}$^*$) \\
\hline
$Q_i>Q_0$ & & $6.2^{+0.2}_{-0.3}$ & $5.6+0.7^*=6.3$ \\
\hline
Mixed & $A,S,T$ & $3.8(1.5)$ \\
 & OPE   & $10.9(0.8)$ \\
 & Eff.\ pole & $1.2$ \\
 & Sum & $15.9(1.7)$ & $11.6+1.9^*=13.5$ \\
\hline
$Q_i<Q_0$ & $A=f_1,f_1',a_1$ & $12.2(4.3)$ & $13.1$ \\
    & $S=f_0,
    a_0$
    & $-0.7(4)$ & \\
    & $T=f_2,a_2$ & $-2.5(8)$ &  $2.9^*$ \\ 
    & Other & $2.0$ & $2.4+5.6^*=8.0$\\
    & Sum & $11.0(4.4)$ & $24.0$  \\
\hline
Sum & & $33.2(4.7)$ &$32.7+11.1^*=43.8$ \\
\hline
\end{tabular}
\end{center}
\renewcommand{\arraystretch}{1.0}
\end{table}

\begin{figure*}[t]
\caption{HLbL summary plot of WP25 \cite{\WPnew} with hQCD prediction [Eq.~(5.52) in WP25] added (red).
The averages are from WP20~\cite{Aoyama:2020ynm} and WP25, respectively, the input used is from the dispersive analysis HSZ-24 \cite{\HSZ} and the lattice-QCD studies of RBC/UKQCD-19~\cite{Blum:2019ugy}, Mainz/CLS-21+22~\cite{Chao:2021tvp,Chao:2022xzg}, RBC/UKQCD-23~\cite{Blum:2023vlm}, and BMW-24~\cite{Fodor:2024jyn}.
(Figure adapted from \cite{\WPnew})}
\label{fig:hlblsummaryplot}       
\centering
\includegraphics[width=0.83\textwidth,clip]{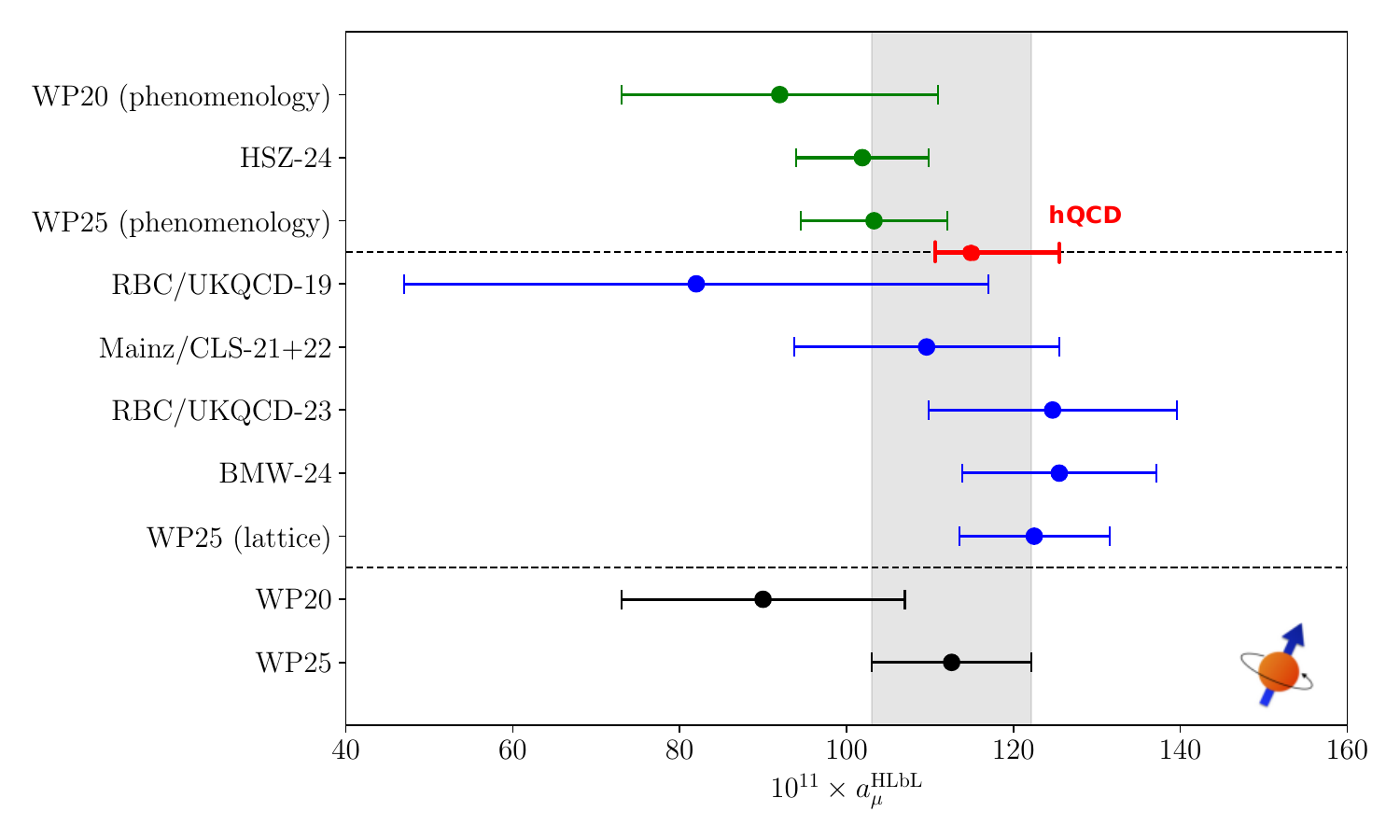}
\end{figure*}

\section{Conclusion}

In the previous White Paper from 2020 \cite{\WPold}, 
the largest
contribution to the uncertainty in the HLbL contribution
was due to the insufficiently determined contribution
of axial-vector mesons and SDCs in the axial sector.
For the new White Paper \cite{\WPnew}, a first
dispersive analysis of these subleading contributions
became available \cite{\HSZ}, with results that
agree remarkably well with the hQCD model result
for the axial sector obtained earlier in \cite{\LMR}.

A new open question appeared, however, concerning the
role of tensor mesons. In the previous White Paper, their
contribution was estimated as \cite{\WPold,Danilkin:2016hnh}
$a_\upmu^\mathrm{T,WP20}=0.9(1)\times 10^{-11}$, 
surprisingly small given the prominence of tensor mesons
in two-photon collisions.
Indeed, this result turned out to require revision; Ref.~\cite{\HSZ} found that 
the same ansatz with a single TFF (out of five possible ones)
in a new formalism that avoids kinematical singularities\footnote{In the new
basis set up in \cite{Hoferichter:2024fsj} these singularities
are only avoided automatically when a subset of the five TFFs is present as is
the case for the quark model and also in hQCD. The results of 
\cite{Hoferichter:2020lap} show, however, that all five TFFs will be
needed in a complete dispersive analysis.}
gives about three times larger results, but with the opposite sign.

In HW hQCD models, tensor mesons have been found by the present
authors to involve a second TFF which contributes only in
the doubly virtual case and which again reverses the sign.
Moreover, the result from summing over all excited tensor mesons
leads to a much larger result than the pole contribution of the
ground-state tensor mesons, $a_\upmu^\mathrm{T,hQCD} \approx 11\times 10^{-11}$.
In hQCD, the tensor sector with its
infinite tower of tensor excitations proves to be important
for the short-distance behavior of the HLbL amplitude beyond the
Melnikov-Vainshtein SDC; the axial sector alone
accounts for only a fraction of the symmetric SDCs, the addition
of tensors closes much of the gap to the pQCD result
(see Table \ref{tab:symmSDCs}).

Without the tensor-sector contribution, the hQCD result of \cite{\LMR} agrees
completely with the current dispersive result for the HLbL contribution,
while its inclusion would remove the tension with the
latest lattice-QCD results which have reached similar
precision as the dispersive approach, see Fig.~\ref{fig:hlblsummaryplot}.
Evidently, it will be important to include a more complete
data-driven analysis of tensor-meson exchange contributions in
the next update of the HLbL contribution to the SM prediction of
the muon $g-2$.

\subsection*{Acknowledgments}
This work was funded in part by the Austrian Science Fund (FWF), grant-DOI \url{https://www.doi.org/10.55776/PAT7221623}.  L.C. acknowledges the support of the INFN research project
ENP (Exploring New Physics).
%
\bibliography{references}

\end{document}